\documentclass[letterpaper,english,aps,prl,floatfix,twocolumn,showpacs,amsfonts,amssymb,superscriptaddress]{revtex4}
\usepackage[T1]{fontenc}
\usepackage[latin1]{inputenc}
\usepackage{amsmath}
\usepackage{babel}
\usepackage{graphics}
\usepackage{amssymb}
\usepackage{amscd}
\usepackage{bm}

\begin{document}

\title{Phase diagram of the anisotropic Kondo chain}
\author{E.~Novais}
\affiliation{Instituto de Física Gleb Wataghin, Unicamp, Caixa Postal 6165, 13083-970
Campinas, SP, Brazil}
\author{E.~Miranda}
\affiliation{Instituto de Física Gleb Wataghin, Unicamp, Caixa Postal 6165, 13083-970
Campinas, SP, Brazil}
\author{A.~H.~Castro~Neto}
\affiliation{Department of Physics, Boston University, Boston, MA 02215}
\author{G.~G.~Cabrera}
\affiliation{Instituto de Física Gleb Wataghin, Unicamp, Caixa Postal 6165, 13083-970
Campinas, SP, Brazil}
\date{\today}
\begin{abstract}
We establish the phase diagram of the one-dimensional anisotropic
Kondo lattice model at $T=0$ using a generalized two-dimensional
classical Coulomb gas description. We analyze the problem by means of
a renormalization group (RG) treatment. We find that the phase diagram
contains regions of paramagnetism, partial and full ferromagnetic
order.
\end{abstract}
\maketitle

The question of the behavior of localized magnetic moments in metals
bears on a variety of important materials, from heavy fermion systems
to manganites. Central to most theoretical studies is the Kondo
lattice model (KLM) with both antiferromagnetic (AFM) and
ferromagnetic (FM) couplings. The former case results from
superexchange and is associated with heavy fermion behavior in
rare-earth and transition metal intermetallic systems
\cite{HeavyFermions}. On the other hand, direct exchange leads to
intratomic FM interactions of the Hund's rule type, also leading to a
KLM description of systems such as the manganites \cite{Dagotto1}. In
both cases, there is strong interest in the determination of the phase
diagram and, more importantly, in the nature of the quantum phase
transitions that separate the various phases at $T=0$ \cite{Doniach}.

\begin{figure}
{\centering\resizebox*{3.0in}{!}{\includegraphics{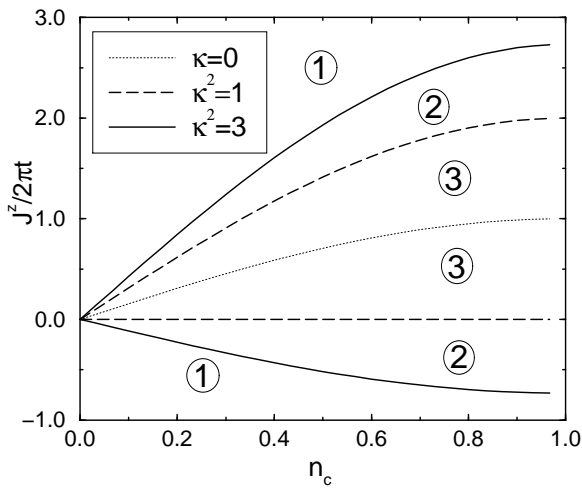}} \par}
\caption{\label{transicaodefase}Ground state phase diagram of the KLM as
a function of the Kondo coupling \protect\( J_{z}\protect \) and
the band filling \protect\( n_{c}\protect \). (1) corresponds to
the ordered phase of the spin array with \protect\( \left\langle
S ^{z}\right\rangle =1/2\protect \), 
(2) is an ordered phase with \protect\( \left\langle
S^{z}\right\rangle <1/2\protect \), and (3) is a paramagnetic phase, 
\protect\( \left\langle S^{z}\right\rangle =0\protect \).}
\end{figure}

Due to the occurrence of non-perturbative Kondo correlations, in
addition to critical order parameter fluctuations, quantum phase
transitions in the AFM KLM are still a hotly debated issue
\cite{piersetal} as compared to their counterparts in other metallic
magnetic systems\cite{Hertz} (see \cite{Si} for a recent attempt). In
this letter, we present a renormalization group (RG) treatment of the
KLM that, although confined to one spatial dimension, has the
advantage of being able to incorporate \textit{both types of
low-energy processes on an equal footing.}  The RG analysis has been
very fruitful in the single-impurity case and is the adequate tool to
address the issue of competing ground states.  Our treatment
generalizes to the lattice case the mapping of the single-impurity
problem into a classical Coulomb gas \cite{Yuval-Anderson3}. This
enables us to decimate both the conduction electrons and the spins
simultaneously.  We establish the phase diagram
(Fig.\ref{transicaodefase}) for both signs of the coupling constant in
a unified fashion. Our results may prove directly useful for
quasi-one-dimensional organic compounds with localized moments, such
as $\rm (DMET)_2FeBr_4$ \cite{enoki}.

The anisotropic KLM chain is described by:
\begin{eqnarray}
H = -t\sum _{j,\sigma }\left( c^{\dagger }_{j+1\sigma }c^{\phantom
    {\dagger }}_{j\sigma }+h.c.\right)+J_{z}S^{z}_{j}s^{z}_{j}
 + J_{\perp }\sum_{\alpha =x,y}
 S_{j}^{\alpha }s_{j}^{\alpha },\nonumber 
\end{eqnarray}
where \( c_{j\sigma } \) annihilates a conduction electron in site \(
j \) with spin projection \( \sigma \), \( \mathbf{S}_{j} \) is a
localized spin \( \frac{1}{2} \) operator and \(
\mathbf{s}_{j}=\frac{1}{2}\sum _{\alpha \beta }c^{\dagger }_{j\alpha
}\bm{\sigma}^{\phantom{\dagger}}_{\alpha\beta} c^{\phantom {\dagger
}}_{j\beta } \), the conduction electron spin density. At long
wavelengths and low energies one can linearize the dispersion around
the Fermi points \( \pm k_{F} \) (\( k_{F}a=\frac{\pi }{2}n_c \), \(
n_c \) is the conduction electron number density), and take the
continuum limit of the fermionic operators.  Following
Ref. \cite{Zachar-Emery-Kivelson} we use bosonization identities and
neglect the backscattering terms, which are irrelevant away from
commensurability. In this limit the bosonic charge fields decouple
giving rise to gapless collective modes.  On the other hand, the
bosonic spin field is coupled to the local spins, leading to the
Hamiltonian:
\begin{eqnarray}
H & = &
\frac{v_{F}}{2}\int dx\left[ \partial _{x}\phi _{c,s}\left( x\right)
 \right] ^{2}+\left[ \partial _{x}\theta _{c,s}\left( x\right) \right] ^{2}+
\label{bosonham} \\
& &
\sum _{i} \left\{ J_{z}a\sqrt{\frac{2}{\pi }}\partial _{x}
\phi _{s}\left( i\right) S^{z}\left( i\right) + \left[
\frac{J_{\bot }a}{2\pi \alpha }\times \right. \right.
\nonumber \\
& &
\left.\left.\times e^{-i\sqrt{2\pi }\theta _{s}\left( i\right) }
\cos \left[ \sqrt{2\pi }\phi _{s}\left( i\right)
 \right] S^{-}\left( i\right) + h.c.\right]\right\} ,
\nonumber
\end{eqnarray}
where $v_F=2t\sin k_Fa$ is the Fermi velocity and the bosonic fields are
defined as in Ref.~\cite{Zachar-Emery-Kivelson}.

As in the Kondo problem, it is convenient to
rescale the Hamiltonian by the Fermi velocity, introducing the
dimensionless coupling constants \( \tilde{J}_{z,\bot
}=\frac{aJ_{z,\bot }}{v_{F}} \). We follow an approach analogous
to the Anderson-Yuval-Haman mapping
of the Kondo impurity problem onto a classical Coulomb gas (CG)
\cite{Yuval-Anderson3,Lee}.  This is achieved by going to a path integral
formulation in the coherent state basis of the bosonic fields and the
\( S^{z} \) basis of the local moments. The \( z \)-part is unmodified
whereas the transverse terms generate spin flips along the Euclidean
time direction leading to the partition function:
\begin{equation}
\label{partitionfuction}
Z=\int D\phi D\theta \sum _{\left\{ \sigma \right\} }
\sum _{v=\pm 1}y^{N}e^{-S}.
\end{equation}
 Here, \( y=\frac{\left| \tilde{J}_{\bot }\right| }{2} \), \( N \)
is the number of flips for a certain
spin configuration \( \{S_z\} \), and the Euclidean action is:
\begin{eqnarray}
S & = &
S_{0}+\tilde{J}_{z}\sqrt{\frac{2}{\pi }}\sum _{x}\int d\tau \partial
_{x}\phi _{s}\left( x,\tau \right) S^{z}\left( x,\tau \right)
+\label{originalaction}\\ 
& & -i\sqrt{2\pi }\sum _{j}\left[v\left( j\right) q\left( j\right) \phi
_{s}\left( j\right) -\theta _{s}\left( j\right) q\left( j\right)\right]
\nonumber , 
\end{eqnarray}
where $S_0$ is the free Gaussian bosonic action in both variables
$\phi_s$ and $\theta_s$\cite{Gogolin}, \( \sum _{j} \) is a sum over
kink (spin flip) coordinates and \( q=S^{z}\left( \tau +\delta
\tau\right) -S^{z}\left( \tau \right) =\pm 1 \), is called the
``magnetic charge''\cite{Gogolin}. Because of the cosine in the
\(\tilde{J}_{\bot } \) term of (\ref{bosonham}), for each spin flip
the \( \phi _{s}\) field comes in with two different signs
\cite{Gogolin}. We denote them by \( v(j)=\pm 1 \) (\( v(j)q(j) \) is
called the {}``electric charge{}''). Thus, each particle corresponds
to a spin flip and has both a magnetic and an electric charge, which
are related to the original term that produced the flip. The fugacity
of the particles is \( y \).  For instance, a spin flip produced by
the right moving fermions corresponds to a particle with electric and
magnetic charges with the same sign, whereas one produced by the left
moving fermions gives rise to a particle with opposite signs on its
charges. There are two restrictions on the charge configurations
\textit{for each space coordinate}: (i) the magnetic charge \( q \)
must alternate along the time direction (because its origin is a
spin-$1/2$ flip) and (ii) the total charges must be zero \( \sum
q=\sum qv=0 \) (because of periodic boundary conditions in the time
direction).  We note that these conditions are more stringent than in
usual 1D bosonic field theories\cite{Gogolin}. We therefore call them
strong neutrality conditions.

The final step consists of tracing out the bosonic fields in
(\ref{originalaction}) in order to obtain an effective action for the
spins and kinks. When this is done, both short and long range
interactions are generated.  The latter are universal but the short
range ones depend on the cutoff procedure\cite{Wiegmann}. These short
range terms are essentially the same found in reference\cite{Gulacsi1,Gulacsi2}
by means of a modified bosonization approach. We will focus on the
universal long range part of the action. Upon integrating by parts in
imaginary time, spin time derivatives become kink variables. 
We rewrite all long range terms in the form of a generalized CG
action in two-dimensional Euclidean space for the kinks
\cite{Niehnus}:
\begin{equation}
\label{coulombgas}
S_{eff}=\begin{array}[t]{l}
N\ln y + \frac{1}{2}\sum _{ij}\left[ \frac{\kappa ^{2}}{g}
\ln \left| r_{ij}\right| m\left( i\right) m\left( j\right) +\right.\\
\\
\left. 
+ g\ln \left| r_{ij}\right| e\left( i\right) e\left( j\right)
 -i\kappa \varphi _{i,j}e\left( i\right) m\left( j\right) \right],
\end{array}
\end{equation}
where $r_{ij}$ is the Euclidean distance between 2 particles and $\varphi _{i,j}$
the angle. In (\ref{coulombgas}), \( \kappa =1-\tilde{J}_{z}/\pi  \),
\( g=1 \), \( m\left( j\right) =q\left( j\right)  \) is the magnetic
charge and \( e\left( j\right) =v\left( j\right) q\left( j\right) \)
is the electric one. The coefficient of the term in \( \varphi _{ij}
\) is usually an integer (the conformal spin, \cite{Gogolin}) and the
ambiguity of \( 2\pi n \) in the angle is then irrelevant. However, in
this case, $\kappa$ can assume non-integer values. The theory remains
well defined nevertheless, due to the strong neutrality condition,
which leads to a cancellation of the Riemann surface index.

In order to investigate the physics of the action~(\ref{coulombgas}),
we employ a renormalization group (RG) procedure\cite{Niehnus}. The
most interesting situation is the dense limit, where the distance
between impurities is of the order of the smallest bosonic wavelength
available in the system. Even though we begin with unitary charges,
higher charges are generated by renormalization, which come from the
fusion of elementary kinks\cite{Yakovenko1}. They correspond to new
action terms with spin flip \textit{pairs} and four fermion operators:
\begin{eqnarray}
O_{ph} & \sim  & \widetilde{G}\left[ \psi _{R\uparrow }^{\dagger }
\left( x\right) \psi _{R\downarrow }\left( x\right)
 \psi _{L\uparrow }^{\dagger }\left( x\right)
 \psi _{L\downarrow }\left( x\right) \times \right. \nonumber \\
 &  & \left. \times S^{-}\left( x+\delta \right) S^{-}\left( x\right) + h.c.
\right] ,\label{oph} \\
O_{pp} & \sim  & G\left[ \psi _{R\uparrow }^{\dagger }\left( x\right)
 \psi _{R\downarrow }\left( x\right)
 \psi _{L\downarrow }^{\dagger }\left( x\right)
 \psi _{L\uparrow }\left( x\right) \times \right. \nonumber \\
 &  & \left. \times S^{+}\left( x+\delta \right) S^{-}\left( x\right) + h.c.
\right] ,\label{opp} 
\end{eqnarray}
where \( \delta \) is a distance of order \( \alpha \). These operators
do not appear in the original Hamiltonian but are generated by the RG
procedure. Notice that they are associated with exchange
processes generated by electron-electron interactions. 
It is natural, from this view point, to think of the localized spins
are generating interactions among the electrons. The \( O_{ph}
\) term flips two nearby spins simultaneously and its action generates
a particle with charges $(m,e)=(\pm 2,0)$, whereas \( O_{pp} \)
creates $(0,\pm 2)$ charges.  Thus, \( G\) and \( \widetilde{G} \) are
the fugacities of these charge 2 particles. Higher charges are
irrelevant. The RG equations are: 
\begin{eqnarray}
\frac{dy}{d\ell } & = & \left( 2-\frac{1}{2}
\left( \frac{\kappa ^{2}}{g}+g\right) \right) y+
\pi \varepsilon y\left(G+\tilde{G}\right) ,
\nonumber 
\\
\frac{d\tilde{G}}{d\ell } & = & 2\left( 1-\frac{\kappa ^{2}}{g}
\right) \tilde{G}+\pi y^{2},
\nonumber 
\\
\frac{dG}{d\ell } & = & 2\left( 1-g\right) G+\pi y^{2},
\nonumber 
\\
\frac{1}{\pi }\frac{d\ln g}{d\ell } & = & \frac{\varepsilon }{2}
\frac{\kappa ^{2}-g^{2}}{g}y^{2}+
\frac{\kappa ^{2}}{g}\tilde{G}^{2}-gG^{2},
\label{rg} 
\end{eqnarray}
 where
$\varepsilon =\sin \left( 2\pi \kappa \right)/(2\pi \kappa),$
 with initial conditions:
$g\left( 0\right) =1,\, y\left( 0\right) =
\tilde{J}_{\bot }/2,\: \textrm{and}\: G\left( 0\right) =\tilde{G}\left( 0\right) =0.$

 The Coulomb coupling \( g \) starts at 1 for non-interacting conduction
electrons. However, the same RG equations apply to the case of conduction
electrons with an SU(2) non-invariant forward scattering interaction.
In this case, the initial value of \( g \) is the corresponding Luttinger
liquid parameter\cite{Gogolin}. We will not consider this
case here, but its phase diagram is analogous to the one below.

By considering the solutions of Eqs.~(\ref{rg}), we can trace three
distinct regions characterized by different fugacity flows.
In Fig.~\ref{transicaodefase} these regions are plotted as a
function of the original Kondo coupling $J_z$ and the band filling $n_c$.
Since the RG equations depend only on \(
\left| \kappa \right| \), those regions are mirror reflections on the
\( \kappa =0 \) line. The full and dashed lines trace out borders
between different phases, whereas the dotted line, which is embedded
in region 3, is the {}``Toulouse line{}'' of
Ref.~\cite{Zachar-Emery-Kivelson}.

In region 1 (\( \kappa ^{2}>3\)), single spin flip processes are
irrelevant (\( y\rightarrow 0 \)) just like the FM phase of the single
impurity Kondo problem\cite{Yuval-Anderson3}. Besides, this
phase has one of the higher charges \( G \) flowing to strong coupling
(see (\ref{opp})). In contrast, both single and double spin flip
processes are relevant in regions 2 and 3. What distinguishes them is
the fact that in region 2 the flow of $y$ is slower and
$G>\widetilde{G}$.  The flow in the dashed line between 2 and 3 can be
solved analytically: \( y(\ell )=y(0)e^{\ell } \) and \(
G(\ell)=\widetilde{G}(\ell)= \pi y^{2}(0)(e^{2\ell}-1)/2 \).  This defines a
characteristic length \( a\sim 2a_{0}\ln \left( 2/\left|
\tilde{J}_{\bot }\right|\right) \) where \( y(a/a_0)\sim 1\). In this
case, there is a precise balance between the electric and magnetic
charges, which prevents them from being screened. As a consequence, \(
g =1 \) and does not renormalize, though the ground state is a
plasma. All correlations fall in a power law fashion implying a
gapless system.  On the other hand, outside the line \(
\left|\kappa\right| =1 \), the interactions are screened (\(g\rightarrow
0{\textrm{ or }}\infty \)).  A particularly simple case of this kind
of flow occurs in the {}``Toulouse line{}'' (\(\kappa =0 \)), where \(
g\rightarrow 0 \) and all fugacities grow.

Although the RG flows are clear, their physical interpretation is less
straightforward. Since we used Abelian bosonization, we treated in
different ways the \( z \) and the transverse components of the spins.
Therefore, while short range transverse spin correlations are
generated by fusion of elementary particles (Eqs.~(\ref{oph}) and
(\ref{opp})), the corresponding \( z\) correlations appear only
through their annihilation and no fugacity is associated with this
process. Nevertheless, we can make progress by writing down the
operators describing this annihilation. After point-splitting the
fermionic part, we get:
\begin{eqnarray}
O_{z} & \sim &
2\left( \tilde{G}^{2}-G^{2}\right) S^{z}\left( x+\delta \right)
S^{z}\left( x\right) +\nonumber \\ 
& & \frac{y^{2}}{2}S^{-}\left( x+\delta \right) S^{+}\left( x\right) +h.c.,
\end{eqnarray}
which are the counterparts of the transverse terms coming from the
fusion of particles. This enables us to determine the magnetic phase
diagram assigning an effective spin Hamiltonian to some special cases.
Taking $\delta$ to be the lattice spacing at the final RG scale in the
\(O_{pp}\), \(O_{ph}\) and \(O_z\) definitions, we find an effective
Hamiltonian
\begin{eqnarray}
H_{eff} & \sim & H_{0}\left( \bar{\phi_s} ,\bar{\theta_s}\right) +
\sum _{j}2\left[ \tilde{G}^{2}-G^{2}\right]
S^{z}\left( j+1\right) S^{z}\left( j\right) \nonumber \\
& & +y\cos \left( \sqrt{2\pi g}\bar{\phi_s} \left( j\right) \right)
e^{-i\sqrt{\frac{2\pi }{g}}\kappa \bar{\theta_s} \left( j\right) }
S^{-}\left( j\right) \nonumber \\
& & +\left[ G\cos \left( \sqrt{8\pi g}\bar{\phi_s} \left( j\right) \right)
+\frac{y^{2}}{2}\right] S^{-}\left( j+1\right) S^{+}\left( j\right)\nonumber \\
& & +\tilde{G}e^{i\sqrt{\frac{8\pi }{g}}\kappa \bar{\theta_s} \left( j\right) }
S^{-}\left( j+1\right) S^{-}\left( j\right) \: +\: h.c.
\label{hameff}
\end{eqnarray}
It reproduces the same CG that we studied above.  In region 1, this
reduces to the FM Heisenberg model in its ordered phase ($G \sim 1,
\left\langle S^{z}\right\rangle =1/2 $). The effective Hamiltonian for
the \( \kappa =0 \) line is also independent of the bosonic field. It
is an AFM XYZ model in an external field. In this case, the effective
spin Hamiltonian is ordered in the XY plane, $G\sim\widetilde{G}\sim
y\sim 1$.  Nevertheless, this does not imply any order of the {\em
original} spins. A unitary transformation connects the spins of
Eq.~(\ref{hameff}) to the ones of the original model~(\ref{bosonham})
ensuring that the latter are disordered even if the former are ordered
(see the discussion after Eq.~(\ref{rotation}) and
Ref.~\cite{Zachar-Emery-Kivelson}). Another situation that can be
insightful is the \( \left| \kappa \right| =1 \) line. In this case,
we cannot write an effective model for the spins independent of the
bosonic field. However, due to the symmetric flow of \( G \) and \(
\tilde{G}\), the $z$-term vanishes and the order parameter \(
\left\langle S^{x,y,z}\right\rangle \) is still zero.  With these
assignments in mind, we propose that the entire region 3 is a
paramagnetic phase with short-range AFM fluctuations.  There is no
simple effective Hamiltonian within region 2, but the disordering
term, proportional to \( y \), starts to grow more slowly and the
short range \( z \) correlations turn from antiferro- to
ferromagnetic. We thus find that this is an ordered phase with
unsaturated magnetization of the spins.

The picture that emerges from these results is that there are two
continuous phase transitions in the KLM. The first transition from
region 1 to region 2 in Fig.\ref{transicaodefase}, reminiscent of the
Berezinskii-Kosterlitz-Thouless transition of the single impurity
Kondo model \cite{Yuval-Anderson3}, separates regions of relevance and
irrelevance of the single flip process.  The effective model
(Eq.~\ref{hameff}) in region 1 has FM order, with full saturation of
the \textit{localized} spins. A regime with FM order is beyond the
present bosonization treatment, since the spin polarization of the
conduction electrons must be incorporated.  However, the RG flow is
still able to indicate its existence through the irrelevance of single
spin flips.  In a highly anisotropic model, this leads to the ordering
of the localized spin array so that the electrons can gain kinetic
energy (resembling the double exchange mechanism).  In
Refs.~\cite{Sigrist1,Tsunetsugu}, the authors showed that in the
isotropic case this is indeed what happens.  Within this scenario, the
total spin per site (electrons+spins) would be
$S_{tot}^{z}=\left\langle S^{z}\right\rangle-n_{c}/2=(1-n_{c})/2$ in
the AFM case and $S_{tot}^{z}=(1+n_c)/2$ in the FM one.  Within the
region where the couplings are relevant, there is another continuous
phase transition from region 2 to region 3 in
Fig.\ref{transicaodefase}, similar to the transition of the Ising
model in a transverse field
\cite{Kogut1}, that separates a paramagnetic phase (region 3 of
Fig.~\ref{transicaodefase}) from a region with unsaturated
magnetization of localized spins, which grows continuously until the
border of the first transition (region 2 of
Fig.~\ref{transicaodefase}).  This interpretation is consistent with
the numerical studies of both the isotropic FM KLM of Dagotto
\textit{et al.} \cite{Dagotto1} and the isotropic AFM KLM of
Tsunetsugu\textit{et al.} \cite{Tsunetsugu}. The methods used here
cannot describe the region of phase separation found in the numerical
simulations for the FM KLM, because in that case the magnetic energies
are of the order of the electron bandwidth. In this limit the
bosonization scheme is not applicable.

We now make contact with previous treatments of the KLM with Abelian
bosonization.  In Refs.~\cite{Zachar-Emery-Kivelson,Gulacsi1,Gulacsi2}, a
family of unitary transformations,
\begin{eqnarray}
U=e^{-if(J_{z}) \sum_{i} \theta_s(i)S^{z}(i)},
\label{rotation}
\end{eqnarray}
is used to define new fields that mix spin and boson degrees of
freedom.  The charges of the CG we obtained arise from vortices of
these mixed fields.  As can be readily checked, the integration by
parts that we performed in the effective action for spins and kinks is
equivalent to this unitary transformation.  Therefore, our
effective Hamiltonians in the different regions of RG flow (Eq.~\ref{hameff})
should be understood \textit{in this rotated basis}. This is of special
importance in the analysis of the $\kappa=0$ line, where
$\langle\theta_s\rangle=0$. Thus, even if the spins acquire order in the
XY plane, they still remain disordered in the original basis. Zachar
\textit{et al.} \cite{Zachar-Emery-Kivelson} argued that along the
$\kappa=0$   line the   system   has a spin gap. Since   the   effective
Hamiltonian~(\ref{hameff}) in this line is in a gapped phase,
our results are consistent with this conclusion.
However, this  is at variance with the available
numerical   evidence\cite{Tsunetsugu,Sikkema2} for the {\em isotropic}
KLM.    This  discrepancy  raises   the  question   about  whether the
anisotropic model can  capture the physics of  the isotropic one.   In
addition,  the   subsequent   work of  Zachar~\cite{Zachar}   proposes
additional phases  away from $\kappa=0$, which  may be  related to our
regions 1 and  2. Honner  and Gul\'{a}csi\cite{Gulacsi1,Gulacsi2} have
also proposed a phase diagram for  the \textit{isotropic} 1D KLM. They
predict  a paramagnetic phase for  ferromagnetic  coupling. This is in
disagreement with  our  results and  the  work of   Dagotto \textit{et
al}\cite{Dagotto1}. A full discussion of  their methods and results in
contrast to ours will be published elsewhere.

In conclusion, we have established the zero temperature phase diagram
of the anisotropic 1D KLM with ferromagnetic and antiferromagnetic
coupling. We have found three different phases: a paramagnetic phase
where the Kondo effect dominates; a fully polarized magnetic phase
where the ``double exchange'' correlations drive the system towards
order; and a partially polarized phase where Kondo effect and magnetic
correlations compete directly to generate partial polarization.  The
two quantum phase transitions have continuous nature, closely related
to the Berezinskii-Kosterlitz-Thouless transition of the single
impurity Kondo problem and to the Ising model in a transverse field.
Although we have worked in 1D, many of the effects discussed here are
generic and also occur in higher dimensions. In spite of the fact that
we have used Abelian bosonization and worked on the anisotropic model,
our findings are in agreement with the numerical simulations in the
SU(2) KLM \cite{Dagotto1,Tsunetsugu}. It would be interesting to have
the numerical work extended to the anisotropic model as a further test
of our results. Finally, we hope our results will be a stimulus for
the study of the phase diagram of quasi-one-dimensional systems with
localized moments such as $\rm (DMET)_2FeBr_4$ \cite{enoki}.

We thank I.~Affleck, A.~L.~Chernyshev, E.~Dagotto, M.~Gul\'acsi, 
N.~Hasselmann, S.~Kivelson, S.~Sachdev and O.~Zachar, for suggestions
and discussions. 
This work was supported by FAPESP and CAPES (Brazil).
A.~H.~C.~N. acknowledges 
partial support provided by a CULAR grant under the auspices
of the US DOE.


\end{document}